\documentclass[10pt,twocolumn,reprint,amsmath,amssymb,aps,prb,showpacs,longbibliography,superscriptaddress]{revtex4-1}
\usepackage[latin9]{inputenc}
\usepackage{amsmath}
\usepackage{amssymb}
\usepackage{graphicx}
\usepackage{esint}

\makeatletter
\usepackage{amsfonts}
\usepackage{graphics}

\makeatother

\begin{document}
\title{Microscopic many-body theory of two-dimensional coherent spectroscopy
of excitons and trions in atomically thin transition metal dichalcogenides}
\author{Hui Hu}
\affiliation{Centre for Quantum Technology Theory, Swinburne University of Technology,
Melbourne 3122, Australia}
\author{Jia Wang}
\affiliation{Centre for Quantum Technology Theory, Swinburne University of Technology,
Melbourne 3122, Australia}
\author{Xia-Ji Liu}
\affiliation{Centre for Quantum Technology Theory, Swinburne University of Technology,
Melbourne 3122, Australia}
\date{\today}
\begin{abstract}
We present a microscopic many-body theory of the recently measured
two-dimensional coherent spectroscopy (2DCS) of excitons and trions
in monolayer MoSe$_{2}$ materials {[}K. Hao \textit{et al.}, Nano
Lett. \textbf{16}, 5109 (2016){]}, where excitons and trions can be
well interpreted as repulsive and attractive polarons, respectively,
in the dilute limit of exciton density. We derive a simple relation
for the 2DCS spectrum in terms of a modified, mixing time-dependent
polaron Green function, which is valid in the single exciton limit.
Our simulated spectra are in excellent qualitative agreement with
experiments without introducing any phenomenological parameters such
as decoherence rates. In particular, quantum beats between the off-diagonal
crosspeaks in the experimental 2DCS spectra are well reproduced. Our
work, therefore, clarifies the microscopic principle that underlies
the observed optical signals of exciton-trion coherence. We find that
there are two quantitative discrepancies between theory and experiment:
the smaller than expected crosspeak strength and the slightly unsynchronized
quantum beats at different crosspeaks. Tentatively, we attribute these
residual discrepancies to the finite exciton density and the resultant
polaron-polaron interaction, which is not taken into account in our
theory.
\end{abstract}
\maketitle

\section{Introduction}

Over the last decade, atomically thin transition metal dichalcogenides
(TMD) have received increasing attention \citep{Novoselov2005,Wang2018,Berkelbach2018}
due to their unique physical properties arising from extreme low-dimensional
constraints. These two-dimensional (2D) materials are expected to
be promising candidates for a wide range of applications in ultrathin
low-power electronics, optoelectronics, and spintronics. For this
perspective, different types of experimental spectroscopy techniques
have been used to characterize optical properties of monolayer TMD
materials \citep{Wang2018,Berkelbach2018}, including the reflection
or absorption spectra \citep{Mak2010}, photoluminescence \citep{Gutierrez2013},
nonlinear two-pulse pump-probe measurement \citep{Ahmed2020,Tan2020},
and nonlinear four-wave-mixing \citep{Hao2016NatPhys}.

Here, we are specifically interested in the nonlinear two-dimensional
coherent spectroscopy (2DCS) based on the four-wave-mixing \citep{Jonas2003,Li2006,Cho2008},
which enables the study of excited-state dynamics on femtosecond (fs)
timescales and maps out the full third-order nonlinear optical susceptibility
of 2D materials by correlating excitation and emission energies \citep{Cho2008}.
2DCS has been applied to probe the formation and dynamics of excitons
and higher-order excitonic complexes such as trions and bi-excitons
in both molybdenum-based (Mo$X_{2}$) and tungsten-based (W$X_{2}$)
TMD materials \citep{Hao2016NatPhys,Hao2016NanoLett,Hao2017,Muir2022}.
A remarkable recent experimental discovery is the quantum coherence
between trions and excitons in monolayer MoSe$_{2}$, as revealed
by quantum beats between the two off-diagonal crosspeaks at the timescale
of 100 fs \citep{Hao2016NanoLett}. However, due to the lack of theoretical
interpretation of 2DCS spectrum at the microscopic level \citep{Li2006},
it remains a challenge to clarify the microscopic mechanisms underlying
such quantum beats.

In this respect, two pioneering theoretical analyses are worth mentioning
\citep{Tempelaar2019,Lindoy2022}. One is the combined use of the
perturbative Fermi golden rule and the few-body solution for excitons
and trions by Tempelaar and Berkelbach \citep{Tempelaar2019}. As
in the experiment, the electron density could be nonzero, a trion
is now commonly viewed as an attractive polaron \citep{Sidler2017,Efimkin2017},
i.e., a quasiparticle formed by dressing an exciton with particle-hole
excitations of the electron Fermi sea \citep{Massignan2014,Schmidt2018,Wang2022PRL,Wang2022PRA}.
Therefore, in the numerical calculations for three-body trions \citep{Tempelaar2019},
the Brillouin zone sampling resolution has been varied as a way to
effectively tune the electron density and to provide an approximate
polaron description for excitons and trions. In another theoretical
analysis by Lindoy, Chang, and Reichman \citep{Lindoy2022}, which
was posted most recently, the unrealistic limit of an infinitely heavy
hole has been taken in order to utilize the exact solution of the
well-known Mahan-Nozières-De Dominicis (MND) model \citep{Mahan1967,Nozieres1969,Mahan2008}.
However, in the immobile heavy hole limit, polaron quasiparticle resonance
turns into a power-law singularity due to the famous Anderson Anderson\textquoteright s
orthogonality catastrophe \citep{Anderson1967,Knap2012}. Although
the MND model provides useful insight into quantum beats, it is desirable
to consider mobile holes and excitons with finite mass.

In this work, we would like to remove the downsides of these two theoretical
analyses by taking an \emph{exact} polaron model for \emph{mobile}
excitons and trions, with a realistic exciton mass. We present a full
microscopic many-body calculation of the 2DCS spectrum of excitons
and trions. Intriguingly, as photons in four-wave-mixing pulses have
negligible momentum \citep{Mahan2008}, any intermediate non-exciton
states that involve particle-hole excitations of the Fermi sea will
not make contributions to the 2DCS signal due to their different linear
momentum from the initial configuration of the electron Fermi sea.
This is true if we always restrict the maximum number of exciton during
excitations to one in the low exciton density limit. The absence of
particle-hole excitations allows us to derive a simple expression
for the 2DCS spectrum, which provides a microscopic understanding
of the perturbative Fermi golden rule adopted by Tempelaar and Berkelbach
\citep{Tempelaar2019}. The latter was adopted without explanation.

We perform a numerical simulation of the 2DCS spectrum under the experimental
conditions without introducing any phenomenological parameters. There
are excellent qualitative agreements between our theory and the recent
experiment by Hao \textsl{et al.} \citep{Hao2016NanoLett}, indicating
that the microscopic mechanism of quantum beats is indeed captured
by the exciton-trion-polaron model \citep{Sidler2017,Efimkin2017}.
We also find some residual discrepancies, such as the smaller than
expected crosspeak strength and the slightly unsynchronized oscillations
at different crosspeaks. These discrepancies could be due to the polaron-polaron
interaction at finite exciton density, which is not considered in
our calculations but is worth exploring in future works.

It is interesting to note that a cold-atom analog of the 2DCS spectroscopy
was recently proposed by us (i.e., the 2D Ramsey spectroscopy) \citep{Wang2022arXiv1,Wang2022arXiv2},
in which the role of an exciton is played by a spin-1/2 impurity atom,
and the four-wave-mixing is implemented by using a sequence of Ramsey
$\pi/2$ radio-frequency (rf) pulses to flip the pseudospin of the
impurity \citep{Knap2012}. In this work, we will also briefly compare
the two different two-dimensional spectroscopy.

The rest of the paper is organized as follows. In the following section
(Sec. II), we outline the model Hamiltonian for the exciton-trion-polaron
in 2D materials. In Sec. III, we present the many-body theory of the
2DCS spectroscopy and make a brief comparison to the 2D Ramsey spectroscopy
with cold-atoms \citep{Wang2022arXiv2}. In Sec. IV, we first discuss
the details of our numerical calculations and then show the theoretical
results in comparison with the experimental data \citep{Hao2016NanoLett}.
Finally, Sec. V is devoted to conclusions and outlooks.

\section{Model Hamiltonian}

In monolayer MoSe$_{2}$, spin-up (spin-down) electrons and holes
near the $K$ ($K'$) valley can form tightly bound excitons, with
binding energy about several hundred meV \citep{Wang2018,Hao2016NanoLett}.
These excitons can also attract extra spin-opposite electrons (to
be described by the creation and annihilation field operators $c_{\mathbf{k}}^{\dagger}$
and $c_{\mathbf{k}}$) in other valley to form singlet trions, with
a trion binding energy $E_{T}\sim30$ meV \citep{Tempelaar2019}.
In general, the density of extra electrons in other valley is finite,
as characterized by an electron Fermi energy about several ten meV,
$\varepsilon_{F}\sim10$ meV. As the exciton binding energy is significantly
larger than other energy scales in the system, the internal structure
of excitons is frozen and we can describe them by using the creation
and annihilation field operators $X_{\mathbf{k}}^{\dagger}$ and $X_{\mathbf{k}}$.
In the dilute limit of exciton density, the system under consideration
therefore can be well described by a polaron model Hamiltonian ($\hbar=1$)
\citep{Sidler2017,Efimkin2017},
\begin{equation}
\mathcal{H}=\sum_{\mathbf{k}}\left[\epsilon_{\mathbf{k}}c_{\mathbf{k}}^{\dagger}c_{\mathbf{k}}+\epsilon_{\mathbf{k}}^{X}X_{\mathbf{k}}^{\dagger}X_{\mathbf{k}}\right]+U\sum_{\mathbf{qkp}}X_{\mathbf{k}}^{\dagger}c_{\mathbf{q-k}}^{\dagger}c_{\mathbf{q}-\mathbf{p}}X_{\mathbf{p}},\label{eq:Hami}
\end{equation}
where the maximum number of exciton is restricted to $1$, i.e., $\sum_{\mathbf{k}}X_{\mathbf{k}}^{\dagger}X_{\mathbf{k}}\leq1$,
and the density of the electrons ($n=\sum_{\mathbf{k}}c_{\mathbf{k}}^{\dagger}c_{\mathbf{k}}$)
can be tuned by the Fermi energy $\varepsilon_{F}$ through gate voltage.
$\epsilon_{\mathbf{k}}=k^{2}/(2m_{e})$ and $\epsilon_{\mathbf{k}}^{X}=k^{2}/(2m_{X})$
are the single-particle energy dispersion relation of electrons and
excitons, respectively, with electron mass $m_{e}$ and exciton mass
$m_{X}\simeq2m_{e}$ in 2D TMD materials \citep{Wang2018}.

In the dilute limit of electron density ($n\rightarrow0$), the formation
of a trion is driven by the interaction Hamiltonian (i.e., the $U$-term).
Hence, the interaction strength $U$ can be determined by reproducing
the trion binding energy $E_{T}$ \citep{Sidler2017,Efimkin2017}.
In the general case of a finite electron density, it is now understood
that trions and excitons could be well interpreted as the attractive
polarons and repulsive polarons, the two types of quasiparticles that
have been systematically studied over the past fifteen years in ultracold
atomic physics \citep{Massignan2014,Schmidt2018,Wang2022PRL,Wang2022PRA}. 

\section{Many-body theory of two-dimensional coherent spectroscopy}

\begin{figure}
\centering{}\includegraphics[width=0.45\textwidth]{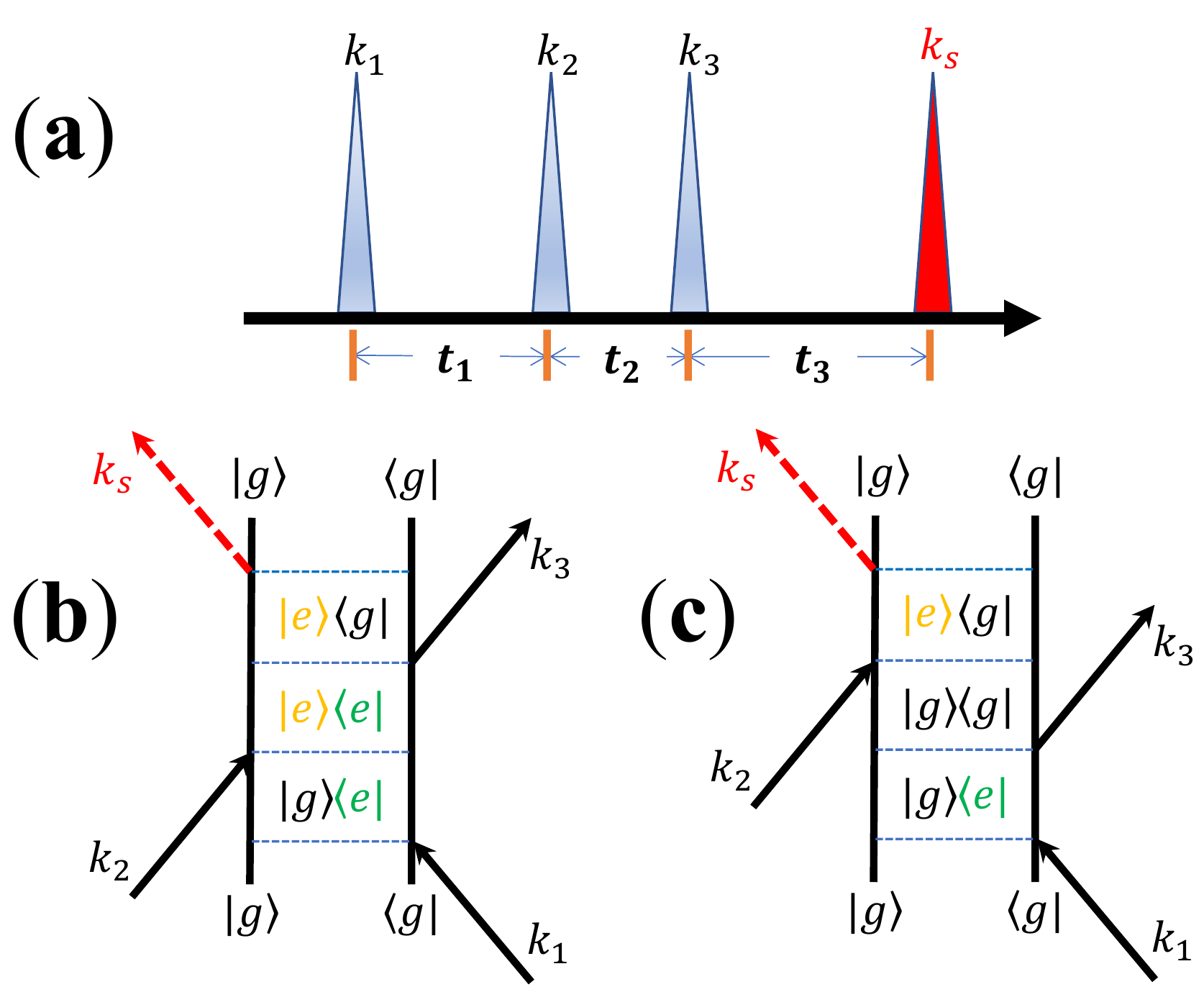}\caption{\label{fig:fig1_sketch}(a) Time ordering of excitation pulses for
standard rephasing 2D coherent spectra. The evolution, mixing, and
emission time delays are labeled as $t_{1}$, $t_{2}$, and $t_{3}$,
respectively. (b) and (c): The two double-sided Feynman diagrams representing
the two contributions to the phase-match directions, $\mathbf{k}_{s}=-\mathbf{k}_{1}+\mathbf{k}_{2}+\mathbf{k}_{3}$.
(b) describes the process of excited-state emission (ESE), $R_{2}(t_{1},t_{2},t_{3})$,
while (c) is referred to as ground-state bleaching (GSB), $R_{3}(t_{1},t_{2},t_{3})$.
Here, we use $\left|g\right\rangle $ and $\left|e\right\rangle $
to denote the many-body states without and with excitons, respectively.
There could be infinitely many excited many-body (polaron) states
$\left|e\right\rangle $, as indicated by different colors. In contrast,
the many-body state of the electron gas $\left|g\right\rangle $ (describing
the Fermi sea) does not change, due to the negligible momentum of
photons in the excitation pulses, as discussed in the text.}
\end{figure}

In the 2DCS spectroscopy \citep{Jonas2003,Li2006}, three pulses are
applied to the sample at times $\tau_{1}$, $\tau_{2}$ and $\tau_{3}$,
respectively, separated by an evolution time delay $t_{1}=\tau{}_{2}-\tau_{1}$
and a mixing time delay $t_{2}=\tau_{3}-\tau_{2}$, as illustrated
in Fig. \ref{fig:fig1_sketch}(a). The nonlinear third-order four-wave-mixing
signal (i.e., the red wave-packet in the figure), as a result of the
three pulses, is measured with frequency-domain heterodyne detection
at a later time $\tau_{s}$, separated from the third pulse by an
emission time delay $t_{3}=\tau_{s}-\tau_{3}$. In the 2DCS experiment
for excitons and trions in MoSe$_{2}$ \citep{Hao2016NanoLett}, the
three excitation pulses and detected signal are all co-circularly
polarized (i.e., $\sigma^{+}$ polarization). As a consequence, only
$K$-valley excitons are created or annihilated by each excitation
pulse, as described by the light-matter interaction operator $\hat{V}$,
\begin{equation}
\sum_{\mathbf{k}}\left(\phi_{\mathbf{k}}e_{\mathbf{k}}h_{-\mathbf{k}}+\phi_{\mathbf{k}}^{*}h_{-\mathbf{k}}^{\dagger}e_{\mathbf{k}}^{\dagger}\right)\propto X_{0}+X_{0}^{\dagger}\equiv\hat{V},
\end{equation}
where $\phi_{\mathbf{k}}$ is the dipole matrix element. As we mentioned
earlier, the proportionality in the above equation (i.e., the introduction
of the exciton field operator) is reasonable in the limit of extremely
large exciton binding energy. We note that, only the zero-momentum
exciton field operators appear in the interaction operator $\hat{V}$,
due to the negligible photon momentum in the excitation pulses (i.e.,
$\mathbf{k}_{i}\sim\mathbf{k}_{s}\sim0$). After some of the three
pulses, different many-body polaron states $\left|e\right\rangle $
with a single exciton are created. These include the inter-valley
trion or attractive polaron, consisting of a $K$-valley exciton and
a $K'$-valley electron, dressed by the multiple-particle-hole excitations
of the electron Fermi sea in the $K'$-valley \citep{Hao2016NanoLett,Tempelaar2019}.

According to the standard nonlinear response theory \citep{Cho2008,Mukamel1995,Zhang2008},
the four-wave-mixing signal is given by the third-order nonlinear
response function, 
\begin{equation}
\mathcal{R}^{(3)}\propto\left\langle \left[\left[\left[\hat{V}\left(t_{1}+t_{2}+t_{3}\right),\hat{V}\left(t_{1}+t_{2}\right)\right],\hat{V}\left(t_{1}\right)\right],\hat{V}\right]\right\rangle ,
\end{equation}
where the time-dependent interaction operator $\hat{V}(t)\equiv e^{i\mathcal{H}t}\hat{V}e^{-i\mathcal{H}t}$,
and $\left\langle \cdots\right\rangle $ stands for the quantum average
over the initial many-body configuration of the system without excitation
pulses, which can be either the ground state at zero temperature or
a thermal state at nonzero temperature. By expanding the three bosonic
commutators, there are a total of four distinct correlation functions
and their complex conjugates \citep{Cho2008}. For the rephasing scheme
(i.e., $t_{1}>0$) with the phase-match directions, $\mathbf{k}_{s}=-\mathbf{k}_{1}+\mathbf{k}_{2}+\mathbf{k}_{3}$,
as adopted in the experiment \citep{Hao2016NanoLett}, we can consider
two significant contributions within the rotating-wave-approximation
in the low exciton density limit, i.e., the process of so-called excited-state
emission (ESE) \citep{Hao2017,Zhang2008}, 
\begin{equation}
R_{2}=\left\langle \hat{V}\hat{V}\left(t_{1}+t_{2}\right)\hat{V}\left(t_{1}+t_{2}+t_{3}\right)\hat{V}\left(t_{1}\right)\right\rangle ,
\end{equation}
and the process of ground-state bleaching (GSB) \citep{Hao2017,Zhang2008},
\begin{equation}
R_{3}=\left\langle \hat{V}\hat{V}\left(t_{1}\right)\hat{V}\left(t_{1}+t_{2}+t_{3}\right)\hat{V}\left(t_{1}+t_{2}\right)\right\rangle .
\end{equation}
These two processes can be understood by using double-sided Feynman
diagrams, as shown in Fig. \ref{fig:fig1_sketch}(b) and Fig. \ref{fig:fig1_sketch}(c),
respectively. Note that, at a small exciton density we only include
the excitations involving a \emph{single} exciton and therefore neglect
the third rephasing process of excited-state absorption (ESA) \citep{Hao2017,Zhang2008},
$R_{1}^{*}(t_{1},t_{2},t_{3})$.

The total rephasing 2DCS spectrum can then be obtained by evaluating
the two correlation functions after a double Fourier transformation
\citep{Li2006,Zhang2008},
\begin{equation}
\mathcal{S}\left(\omega_{1},t_{2},\omega_{3}\right)=\mathcal{S}_{2}\left(\omega_{1},t_{2},\omega_{3}\right)+\mathcal{S}_{3}\left(\omega_{1},t_{2},\omega_{3}\right),
\end{equation}
where for $i=2,3$, 
\begin{equation}
\mathcal{S}_{i}=\intop_{0}^{\infty}\intop_{0}^{\infty}dt_{1}dt_{3}e^{i\omega_{1}^{+}t_{1}}e^{i\omega_{3}^{+}t_{3}}R_{i}\left(t_{1},t_{2},t_{3}\right).
\end{equation}
Here, $\omega_{1}$ is the excitation (or absorption) energy and $\omega_{3}$
is the emission energy. As a response function, we have defined $\omega_{1}^{+}=\omega_{1}+i0^{+}$
and $\omega_{3}^{+}=\omega_{3}+i0^{+}$, where the infinitesimal $0^{+}$
is introduced to regularize the Fourier transformation at $t_{1}\rightarrow\infty$
and $t_{3}\rightarrow\infty$.

\subsection{A simple expression of the 2DCS spectrum}

Let us first focus on the ESE process, $R_{2}(t_{1},t_{2},t_{3})$,
as described by Fig. \ref{fig:fig1_sketch}(b). By inserting the expression
of $\hat{V}_{0}$ and explicitly listing the time-dependence of the
operators, we find that,\begin{widetext}
\begin{equation}
R_{2}\left(t_{1},t_{2},t_{3}\right)=\left\langle \textrm{FS}\right|X_{0}e^{i\mathcal{H}_{X}\left(t_{1}+t_{2}\right)}X_{0}^{\dagger}e^{i\mathcal{H}_{0}t_{3}}X_{0}e^{-i\mathcal{H}_{X}\left(t_{2}+t_{3}\right)}X_{0}^{\dagger}\left|\textrm{FS}\right\rangle e^{-iE_{\textrm{FS}}t_{1}},\label{eq:R2}
\end{equation}
\end{widetext}where $\mathcal{H}_{0}$ and $\mathcal{H}_{X}$ denote
the model Hamiltonian $\mathcal{H}$ in the cases of no exciton and
of a single exciton, respectively. We also denote the initial configuration
of the system (without any excitons and with a background energy $E_{\textrm{FS}}$
of the electron Fermi sea) as $\left|\textrm{FS}\right\rangle $,
so we can evaluate
\begin{equation}
e^{-i\mathcal{H}t_{1}}\left|\textrm{FS}\right\rangle =e^{-i\mathcal{H}_{0}t_{1}}\left|\textrm{FS}\right\rangle =e^{-iE_{\textrm{FS}}t_{1}}\left|\textrm{FS}\right\rangle .
\end{equation}
The configuration $\left|\textrm{FS}\right\rangle $ could be a thermal
mixed state at nonzero temperature. 

In the above expression of $R_{2}(t_{1},t_{2},t_{3})$, we may insert
a complete set of many-body states of the electron Fermi sea, just
before or just after $e^{i\mathcal{H}_{0}t_{3}}$. These many-body
states can formally be constructed by creating multiple-particle-hole
excitations out of the Fermi sea, in the form,
\begin{equation}
\left|\vec{\kappa}_{\nu}\right\rangle =\left[\prod_{i=1}^{\nu}c_{\mathbf{k}_{p}^{(i)}}^{\dagger}\prod_{i=1}^{\nu}c_{\mathbf{k}_{h}^{(i)}}\right]\left|\textrm{FS}\right\rangle ,
\end{equation}
where $\nu$ is the number of particle-hole pairs, $\vec{\kappa}_{\nu}\equiv\{\mathbf{k}_{p}^{(1)},\mathbf{k}_{p}^{(2)},\cdots,\mathbf{k}_{p}^{(\nu)};\mathbf{k}_{h}^{(1)},\mathbf{k}_{h}^{(2)},\cdots,\mathbf{k}_{h}^{(\nu)}\}$
collectively indexes the $\nu$ particle momenta ($\mathbf{k}_{p}^{(i)}>k_{F}$)
and hole momenta ($\mathbf{k}_{h}^{(i)}<k_{F}$), where $k_{F}$ is
the Fermi wavevector. The corresponding momentum and energy of the
many-body state can be denoted as $\epsilon_{\vec{\kappa}_{\nu}}$
and $\mathbf{k}_{\vec{\kappa}_{\nu}}$, respectively. In the absence
of any particle-hole excitations, we simply have $\left|\vec{\kappa}_{\nu=0}\right\rangle =\left|\textrm{FS}\right\rangle $
and $\epsilon_{\vec{\kappa}_{\nu}}=E_{\textrm{FS}}$.

By inserting the identity $\sum_{\vec{\kappa}_{\nu}}\left|\vec{\kappa}_{\nu}\right\rangle \left\langle \vec{\kappa}_{\nu}\right|=1$
into Eq. (\ref{eq:R2}), we find that,
\begin{equation}
R_{2}=\sum_{\vec{\kappa}_{\nu}}F_{\vec{\kappa}_{\nu}}^{*}\left(t_{1}+t_{2}\right)F_{\vec{\kappa}_{\nu}}\left(t_{2}+t_{3}\right)e^{i\epsilon_{\vec{\kappa}_{\nu}}t_{3}}e^{-iE_{\textrm{FS}}t_{1}},
\end{equation}
where we have defined the correlation function,
\begin{equation}
F_{\vec{\kappa}_{\nu}}\left(t\right)\equiv\left\langle \vec{\kappa}_{\nu}\right|X_{0}e^{-i\mathcal{H}_{X}t}X_{0}^{\dagger}\left|\textrm{FS}\right\rangle .
\end{equation}
An immediate observation is that, as a result of the momentum conservation,
the many-body states $\left|\vec{\kappa}_{\nu}\right\rangle $ must
have \emph{zero} momentum. Thus, no particle-hole excitations are
allowed in $\left|\vec{\kappa}_{\nu}\right\rangle $ and the only
contribution to $F_{\vec{\kappa}_{\nu}}(t)$ is provided by the unperturbed
Fermi sea, $\left|\vec{\kappa}_{\nu=0}\right\rangle =\left|\textrm{FS}\right\rangle $.

We now see that the correlation function $F_{\vec{\kappa}_{\nu}}(t)$
can be directly expressed in terms of the retarded polaron Green function
at zero momentum $G_{\mathbf{k}=0}(t)=G(t)$, i.e., 
\begin{equation}
F\left(t\right)=iG\left(t\right)e^{-iE_{\textrm{FS}}t},
\end{equation}
where \citep{Mahan2008}
\begin{equation}
G_{\mathbf{k}}\left(t-t'\right)\equiv-i\theta\left(t-t'\right)\left\langle \textrm{FS}\right|\left[X_{\mathbf{k}}\left(t\right),X_{\mathbf{k}}^{\dagger}\left(t'\right)\right]\left|\textrm{FS}\right\rangle .
\end{equation}
Therefore, we obtain a remarkably simple expression for the ESE contribution
$R_{2}$,
\begin{equation}
R_{2}\left(t_{1},t_{2},t_{3}\right)=G^{*}\left(t_{1}+t_{2}\right)G\left(t_{2}+t_{3}\right).\label{eq:R22G}
\end{equation}

It would be useful to write down a formal expression for the retarded
polaron Green function at zero momentum. For this purpose, we recall
that different zero-momentum polaron states (i.e., the $n$-th state
with polaron energy $E^{(n)}$) can be written as \citep{Wang2022arXiv2,Chevy2006},
\begin{equation}
\left|n\right\rangle =\phi_{0}^{(n)}X_{0}^{\dagger}\left|\textrm{FS}\right\rangle +\sum_{\vec{\kappa}_{\nu\geqslant1}}\phi_{\vec{\kappa}_{\nu}}^{(n)}X_{-\mathbf{k}_{\vec{\kappa}_{\nu}}}^{\dagger}\left|\vec{\kappa}_{\nu}\right\rangle ,\label{eq:PolaronAnsatz}
\end{equation}
where the second term describes the dressing of multi-particle-hole
excitations and the first term describes the ability of free propagation
of the exciton, as measured by the polaron residue $Z^{(n)}\equiv\phi_{0}^{(n)*}\phi_{0}^{(n)}$.
By inserting the identity 
\begin{equation}
e^{-i\mathcal{H}_{X}t}=\sum_{n}\left|n\right\rangle e^{-iE^{(n)}t}\left\langle n\right|
\end{equation}
into the expression of the polaron Green function, we find that,
\begin{equation}
G\left(t\right)=-i\sum_{n}Z^{(n)}e^{-i\mathcal{E}^{(n)}t},
\end{equation}
where $\mathcal{E}^{(n)}=E^{(n)}-E_{\textrm{FS}}$ is the polaron
energy measured with respect to the Fermi sea energy. Therefore, the
ESE contribution $R_{2}$ takes the form,
\begin{equation}
R_{2}=\sum_{nm}Z^{(n)}Z^{(m)}e^{i\mathcal{E}^{(n)}t_{1}}e^{i\left[\mathcal{E}^{(n)}-\mathcal{E}^{(m)}\right]t_{2}}e^{-i\mathcal{E}^{(m)}t_{3}}.
\end{equation}
The physics behind this expression may easily be understood from the
ESE process illustrated in Fig. \ref{fig:fig1_sketch}(b). The factor
$Z^{(n)}Z^{(m)}$ or $\phi_{0}^{(n)}\phi_{0}^{(m)*}\phi_{0}^{(m)}\phi_{0}^{(n)*}$
measures the transfer rates between many-body states induced by the
three excitation pulses and the four-wave-mixing signal, while the
three dynamical (time-evolution) phase factors simply show the phases
accumulated during the time delays $t_{1}$, $t_{2}$ and $t_{3}$,
respectively.

After the double Fourier transformation, we obtain
\begin{equation}
\mathcal{S}_{2}=\sum_{nm}\frac{-Z^{(n)}}{\omega_{1}^{+}+\mathcal{E}^{(n)}}e^{i\left[\mathcal{E}^{(n)}-\mathcal{E}^{(m)}\right]t_{2}}\frac{Z^{(m)}}{\omega_{3}^{+}-\mathcal{E}^{(m)}}.\label{eq:S2}
\end{equation}
It is convenient to introduce a modified, $t_{2}$-dependent polaron
Green function in the frequency domain \citep{Mahan2008},
\begin{equation}
G_{R}\left(\omega,t_{2}\right)\equiv\sum_{n}\frac{Z^{(n)}}{\omega+i0^{+}-\mathcal{E}^{(n)}}e^{-i\mathcal{E}^{(n)}t_{2}},
\end{equation}
which reduces to the conventional retarded polaron Green function
$G_{R}(\omega)$ at zero mixing time delay $t_{2}=0$. Then, the ESE
third-order response function $\mathcal{S}_{2}$ can be written as,
\begin{equation}
\mathcal{S}_{2}\left(\omega_{1},t_{2},\omega_{3}\right)=G_{R}^{*}\left(-\omega_{1},t_{2}\right)G_{R}\left(\omega_{3},t_{2}\right).
\end{equation}

The GSB process $R_{3}$ can be analyzed in the exactly same way.
We find the expressions, $R_{3}(t_{1},t_{2},t_{3})=G^{*}(t_{1})G(t_{3})$
and 
\begin{equation}
\mathcal{S}_{3}\left(\omega_{1},t_{2},\omega_{3}\right)=G_{R}^{*}\left(-\omega_{1}\right)G_{R}\left(\omega_{3}\right).
\end{equation}
The absence of the mixing time ($t_{2}$) dependence in the expressions
is easy to understand from Fig. \ref{fig:fig1_sketch}(c): the system
returns to the initial configuration between the second and third
pulses and therefore does not evolve during the mixing time delay.

By adding the two contributions $\mathcal{S}_{2}$ and $\mathcal{S}_{3}$,
$\mathcal{S}=\mathcal{S}_{2}+\mathcal{S}_{3}$, we arrive at one of
the key results of our work, 
\begin{equation}
\mathcal{S}\left(\omega_{1},t_{2},\omega_{3}\right)=\sum_{nm}\frac{Z^{(n)}Z^{(m)}}{\left(-\omega_{1}\right)^{-}-\mathcal{E}^{(n)}}\frac{1+e^{i\left[\mathcal{E}^{(n)}-\mathcal{E}^{(m)}\right]t_{2}}}{\omega_{3}^{+}-\mathcal{E}^{(m)}},\label{eq:3rdResponseS}
\end{equation}
where $(-\omega_{1})^{-}\equiv-\omega_{1}-i0^{+}$. It is readily
seen that the 2DCS spectrum satisfies the relation, 
\begin{equation}
\mathcal{S}\left(\omega_{1},t_{2},\omega_{3}\right)=\mathcal{S}^{*}\left(-\omega_{3},t_{2},-\omega_{1}\right).
\end{equation}
Therefore, the amplitude and the real part of the 2DCS spectrum are
both symmetric, upon the replacements $-\omega_{1}\rightarrow\omega_{3}$
and $-\omega_{3}\rightarrow\omega_{1}$.

We would like to emphasize that the symmetric 2DCS spectrum is rooted
in two facts. First, the excitation pulse creates or annihilates electron-hole
pairs at essentially zero momentum \citep{Mahan2008}. The electron-hole
pairs can be either tightly bound (i.e., excitons considered in this
work) or loosely bound. On the other hand, we must only take into
account one electron-hole pair in the intermediate excited states.
The existence of the pair-pair correlation, for example, the exciton-exciton
scattering will redistribute the exciton momentum and then lead to
the contributions from the excited Fermi sea $\left|\vec{\kappa}_{\nu}\right\rangle $.
In this case, we can no longer write the response function $R_{2}$
and $R_{3}$ into a product of two polaron Green functions, i.e.,
as given in Eq. (\ref{eq:R22G}).

\subsection{Connection to the Fermi golden rule}

At this point, it is useful to contrast our simple expression of the
2DCS spectrum, Eq. (\ref{eq:3rdResponseS}), with the many-body formalism
used by Tempelaar and Berkelbach (TB) \citep{Tempelaar2019},\begin{widetext}
\begin{equation}
\mathcal{S}_{\textrm{TB}}\left(\omega_{1},t_{2},\omega_{3}\right)=-\left(2\pi\right)^{2}\sum_{\alpha\beta}\left|\left\langle \Psi^{i}\right|\hat{V}\left|\Psi^{\alpha}\right\rangle \right|^{2}\left|\left\langle \Psi^{i}\right|\hat{V}\left|\Psi^{\beta}\right\rangle \right|^{2}e^{-\left(i\omega_{\alpha\beta}+\gamma_{\alpha\beta}\right)t_{2}}\Gamma^{*}\left(E^{\alpha}-E^{i}+\omega_{1}\right)\Gamma\left(E^{\beta}-E^{i}-\omega_{3}\right),\label{eq:S2TB}
\end{equation}
\end{widetext}where $\Psi^{i}$ is the initial state, $\Psi^{\alpha}$
($\Psi^{\beta}$) are the excited states with energies $E^{\alpha}$
($E^{\beta}$), $\omega_{\alpha\beta}\equiv E^{\alpha}-E^{\beta}$
are the energy differences, and $\gamma_{\alpha\beta}$ are the associated
decoherence rates. $\Gamma(\omega)=1/(i\omega-\sigma)$ is the complex
lineshape function with $\sigma$ as the line-broadening parameter. 

It is readily seen that the $t_{2}$-independent term (i.e., the $R_{3}$
contribution) in our Eq. (\ref{eq:3rdResponseS}) is absent in the
TB formalism. This is simply because Tempelaar and Berkelbach focused
on the simulated emission signal \citep{Tempelaar2019}, which is
precisely our ESE contribution $\mathcal{S}_{2}$. We can clearly
see that, if we neglect the phenomenological decoherence rates $\gamma_{\alpha\beta}$
and line-broadening parameter $\sigma$, the TB formalism Eq. (\ref{eq:S2TB})
is essentially identical to our Eq. (\ref{eq:S2}), owing to the correspondences
in the indices $\alpha\leftrightarrow n$ and $\beta\leftrightarrow m$,
in the overlaps $Z^{(n)}\leftrightarrow\left|\left\langle \Psi^{i}\right|\hat{V}\left|\Psi^{\alpha}\right\rangle \right|^{2}$
and $Z^{(m)}\leftrightarrow\left|\left\langle \Psi^{i}\right|\hat{V}\left|\Psi^{\beta}\right\rangle \right|^{2}$,
and finally in the energies $\mathcal{E}^{(n)}\leftrightarrow E^{\alpha}-E^{i}$
and $\mathcal{E}^{(m)}\leftrightarrow E^{\beta}-E^{i}$. Thus, our
derivation of the 2DCS spectrum Eq. (\ref{eq:3rdResponseS}) provides
a useful microscopic explanation to the TB formalism Eq. (\ref{eq:S2TB}).

\subsection{2DCS versus 2D Ramsey spectroscopy}

Let us now briefly compare 2DCS with another type of 2D spectroscopy
with ultracold atoms, the so-called 2D Ramsey spectroscopy \citep{Wang2022arXiv1,Wang2022arXiv2},
in which the exciton in 2DCS is replaced by a spin-1/2 impurity atom.
The spin state of the impurity can be controlled by a rf pulse with
a specific phase \citep{Knap2012} and only the spin-up impurity experiences
an interaction potential with the background Fermi sea. As a result,
the two spin-flip operations, given by the Pauli matrices $\hat{s}^{+}=(\sigma_{x}+i\sigma_{y})/2$
and $\hat{s}^{-}=(\sigma_{x}-i\sigma_{y})/2$, roughly correspond
to the exciton creation operator $X^{\dagger}$ and annihilation operator
$X$. A notable difference is that the excitation rf pulse does not
change the external spatial status of the impurity atom, so it might
be understood as an \emph{effective} light-matter interaction operator
$\hat{V}_{\textrm{eff}}\sim\sum_{\mathbf{k}}(X_{\mathbf{k}}+X_{\mathbf{-k}}^{\dagger})$.
Therefore, as the momentum of the impurity atom (i.e., an effective
exciton) is not restricted to zero, for the intermediate many-body
dynamics the excited Fermi sea with multiple-particle-hole excitations
(as described by $\left|\vec{\kappa}_{\nu}\right\rangle $ with $\nu\geqslant1$)
do make contributions to the third-order response function $\mathcal{R}^{\ensuremath{(3)}}$.
The simple expression found for the 2DCS spectrum, Eq. (\ref{eq:3rdResponseS}),
does not hold in the 2D Ramsey spectroscopy. Additional terms that
make the spectroscopy asymmetric (with respect to exchange the excitation
and emission energies) will appear \citep{Wang2022arXiv2}.

Apart from this difference, there are amazing similarities between
the 2DCS and the 2D Ramsey spectroscopy, although in the latter \citep{Wang2022arXiv1,Wang2022arXiv2}
we have used different notations such as ($\tau,T,t$) for various
time delays and ($\omega_{\tau},\omega_{t}$) as the excitation and
emission energies. In both spectroscopes, phase cycling techniques
can be implemented to select the desired pathways. In the rephasing
mode, the ESE term $R_{2}(t_{1},t_{2},t_{3})$ in the 2DCS is exactly
given by the pathway $I_{1}^{*}(\tau,T,t)$ in the 2D Ramsey spectroscopy
\citep{Wang2022arXiv2} and the GSB term $R_{3}(t_{1},t_{2},t_{3})$
corresponds to the pathway $I_{2}^{*}(\tau,T,t)$ \citep{Wang2022arXiv2}.
Finally, the third-order response function $\mathcal{S}(\omega_{1},t_{2},\omega_{3})$
in the 2DCS precisely corresponds to the symmetric 2D Ramsey response
$\mathcal{A}_{s}^{*}(-\omega_{\tau},T,\omega_{t})$ \citep{Wang2022arXiv2}.

\section{Results and discussions}

\subsection{Computation details}

To demonstrate the usefulness of the simple expression Eq. (\ref{eq:3rdResponseS}),
we perform numerical simulations for monolayer MoSe$_{2}$, with the
effective polaron model Hamiltonian in Eq. (\ref{eq:Hami}). To reduce
the numerical workload, we load the system, which consists of $N$
electrons and a single exciton, onto a two-dimensional square lattice
with $L\times L$ sites. The electron density is then given by $n=N/(La)^{2}$,
where $a$ is the lattice spacing and will be set to be unity ($a=1$)
unless specified otherwise. We assume the electrons and the exciton
move on the lattice with hopping strength $t_{c}$ and $t_{d}$, respectively.
Their single-particle energy dispersion relations are 
\begin{eqnarray}
\epsilon_{\mathbf{k}} & = & -2t_{c}[\cos(k_{x})+\cos(k_{y})]\simeq-4t_{c}+\frac{k_{x}^{2}+k_{y}^{2}}{2m_{e}},\\
\epsilon_{\mathbf{k}}^{I} & = & -2t_{d}[\cos(k_{x})+\cos(k_{y})]\simeq-4t_{d}+\frac{k_{x}^{2}+k_{y}^{2}}{2m_{X}},
\end{eqnarray}
where $m_{e}\equiv1/(2t_{c}a^{2})$ and $m_{X}\equiv1/(2t_{d}a^{2})$
in the dilute limit ($n\rightarrow0$) that of interest. We note that,
the momentum $\mathbf{k}$ on the lattice takes the values, 
\begin{equation}
\left(k_{x},k_{y}\right)=\left(\frac{2\pi n_{x}}{L},\frac{2\pi n_{y}}{L}\right),
\end{equation}
with the integers $n_{x},n_{y}=-L/2+1,\cdots-1,0,1,\cdots L/2$.

\begin{figure}
\begin{centering}
\includegraphics[width=0.45\textwidth]{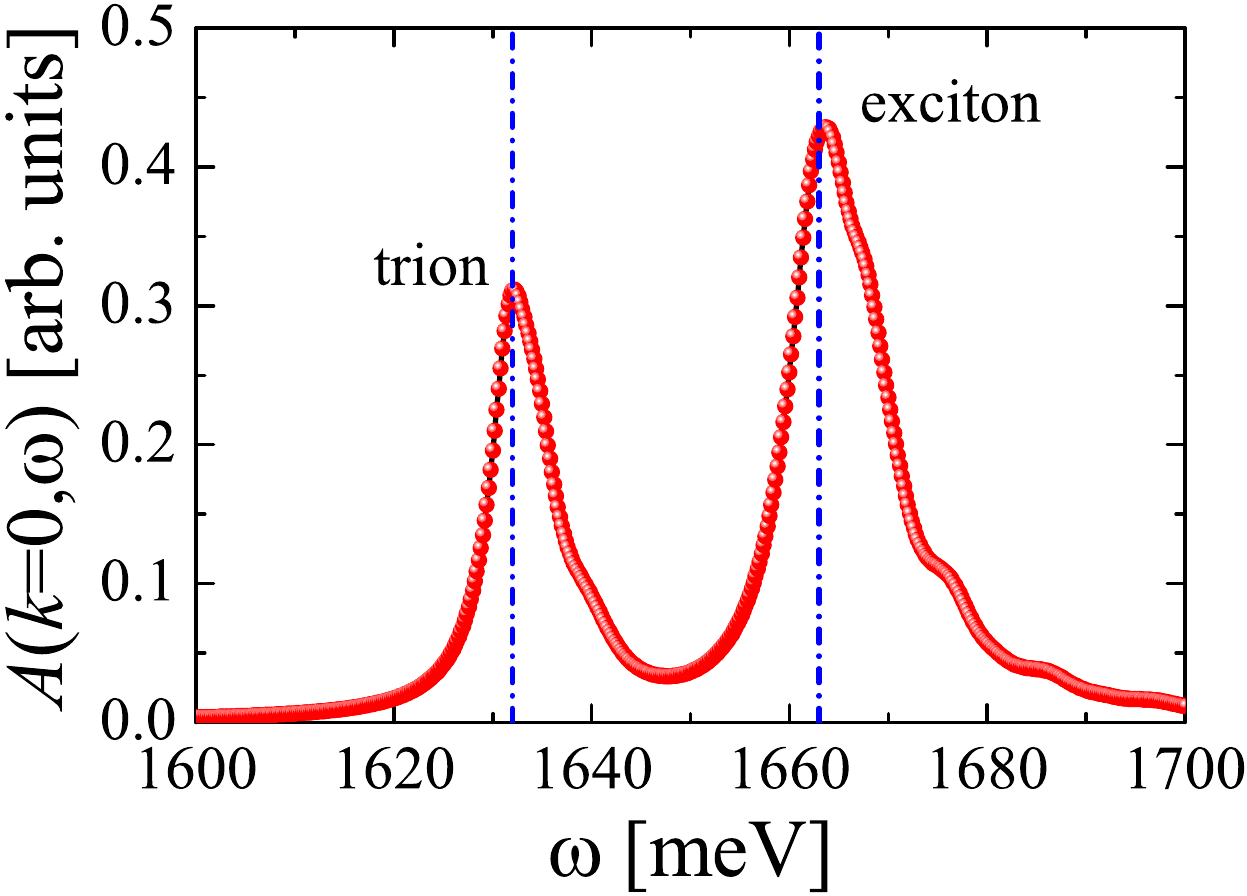}
\par\end{centering}
\centering{}\caption{\label{fig:fig2_akw} The simulated polaron spectral function at the
electron Fermi energy $\varepsilon_{F}=11.8$ meV, revealing the existence
of attractive polaron (trion) and repulsive polaron (exciton). The
two dashed lines indicates the two peak positions for trions (1632
meV) and excitons (1663 meV) shown in the experimental photoluminescence
spectrum of monolayer MoSe$_{2}$ taken at 20 K \citep{Hao2016NanoLett}.}
\end{figure}

We solve the polaron model at zero temperature, by applying the polaron
ansatz Eq. (\ref{eq:PolaronAnsatz}) truncated to one-particle-hole
excitations \citep{Chevy2006}. This is the so-called Chevy ansatz
\citep{Chevy2006}, which is known to yield a quantitatively accurate
prediction for the attractive polaron energy in the strongly interacting
unitary limit \citep{Massignan2014}. In other words, we consider
a Hilbert space constructed by the two kinds of basis states (at zero
polaron momentum), 
\begin{eqnarray}
\left|i\right\rangle  & = & X_{0}^{\dagger}\left|\textrm{FS}\right\rangle ,\\
\left|j\right\rangle  & = & X_{-\mathbf{k}_{p}+\mathbf{k}_{h}}^{\dagger}c_{\mathbf{k}_{p}}^{\dagger}c_{\mathbf{k}_{h}}\left|\textrm{FS}\right\rangle .
\end{eqnarray}
Here, the Fermi sea at zero temperature $\left|\textrm{FS}\right\rangle $
is obtained by filling the single-particle energy level $\epsilon_{\mathbf{k}}$
with $N$ electrons from the bottom of the energy band (i.e., $-4t_{c}$),
up to the energy $E_{F}$. The hole momentum $\mathbf{k}_{h}$ and
the particle momentum $\mathbf{k}_{p}$ must satisfy the constraints
$\epsilon_{\mathbf{k}_{p}}\geqslant E_{F}$ and $\epsilon_{\mathbf{k}_{h}}<E_{F}$,
respectively. The Fermi energy measured from the band bottom is then
\begin{equation}
\varepsilon_{F}=E_{F}+4t_{c}\simeq4\pi nt_{c}=\frac{4\pi N}{L^{2}}t_{c}.
\end{equation}
It is readily seen the dimension of the Hilbert space is $D=1+N(L^{2}-N)$.
Under the basis states, the polaron Hamiltonian then is casted into
a $D$ by $D$ matrix, with the following matrix elements,
\begin{align}
\left\langle \textrm{FS}\right|X_{0}\mathcal{H}X_{0}^{\dagger}\left|\textrm{FS}\right\rangle  & =E_{\textrm{FS}}-4t_{d}+nU,\\
\left\langle \textrm{FS}\right|X_{0}\mathcal{H}X_{-\mathbf{k}_{p}+\mathbf{k}_{h}}^{\dagger}c_{\mathbf{k}_{p}}^{\dagger}c_{\mathbf{k}_{h}}\left|\textrm{FS}\right\rangle  & =U/L^{2},
\end{align}
and\begin{widetext}
\begin{equation}
\left\langle \textrm{FS}\right|c_{\mathbf{k}'_{h}}^{\dagger}c_{\mathbf{k}'_{p}}X_{-\mathbf{k}'_{p}+\mathbf{k}'_{h}}\mathcal{H}X_{-\mathbf{k}_{p}+\mathbf{k}_{h}}^{\dagger}c_{\mathbf{k}_{p}}^{\dagger}c_{\mathbf{k}_{h}}\left|\textrm{FS}\right\rangle =\left[E_{\textrm{FS}}+nU+\epsilon_{\mathbf{k}_{p}}-\epsilon_{\mathbf{k}_{h}}+\epsilon_{-\mathbf{k}_{p}+\mathbf{k}_{h}}^{(I)}\right]\delta_{\mathbf{k}_{p}\mathbf{k}'_{p}}\delta_{\mathbf{k}_{h}\mathbf{k}'_{h}}+\frac{U}{L^{2}}\left(\delta_{\mathbf{k}_{h}\mathbf{k}'_{h}}-\delta_{\mathbf{k}_{p}\mathbf{k}'_{p}}\right).
\end{equation}
\end{widetext}We diagonalize the matrix to obtain the eigenvalues
$E^{(n)}$ and eigenstates, from which we extract the polaron energies
$\mathcal{E}^{(n)}=E^{(n)}-(E_{\textrm{FS}}-4t_{d})$ and the residues
$Z^{(n)}\equiv\phi_{0}^{(n)*}\phi_{0}^{(n)}$. Here, due to the use
of a square lattice, we also need to subtract the lowest energy of
exciton (i.e., $-4t_{d}$) in calculating the polaron energies.

\begin{figure*}
\centering{}\includegraphics[width=0.33\textwidth]{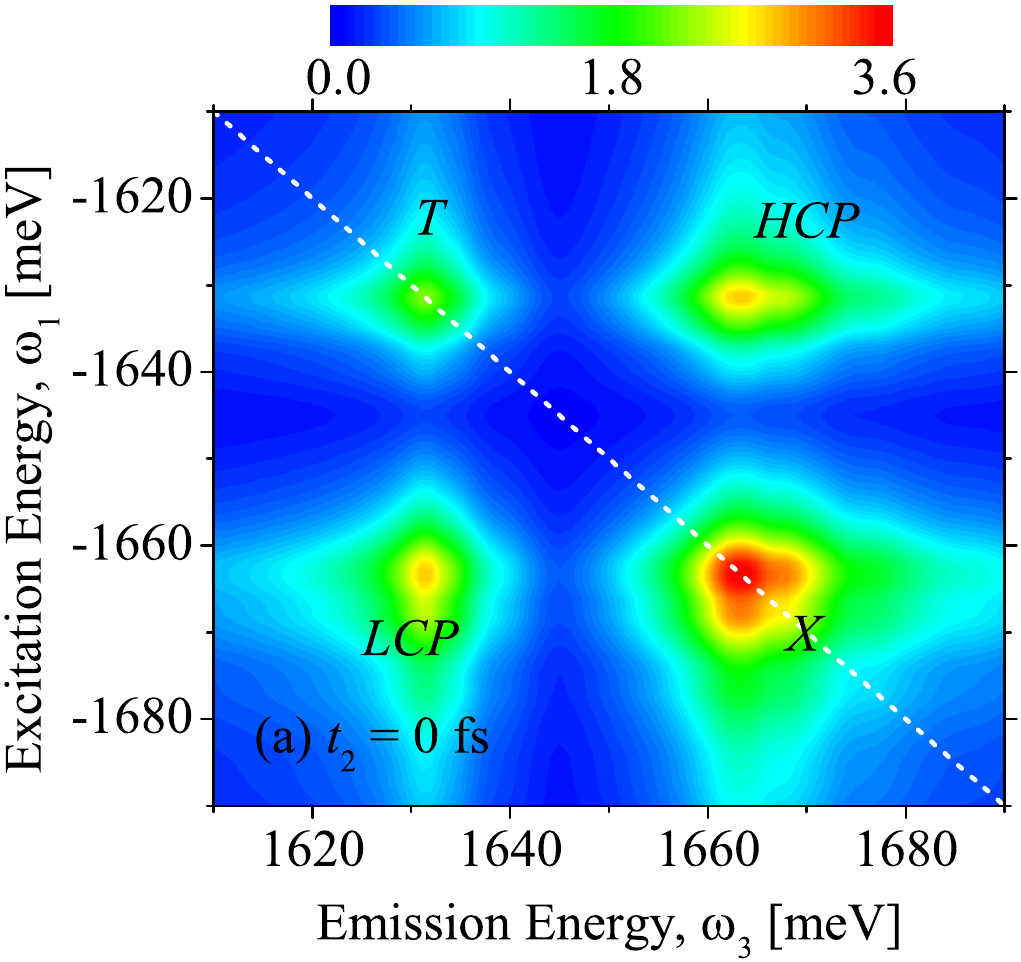}\includegraphics[width=0.33\textwidth]{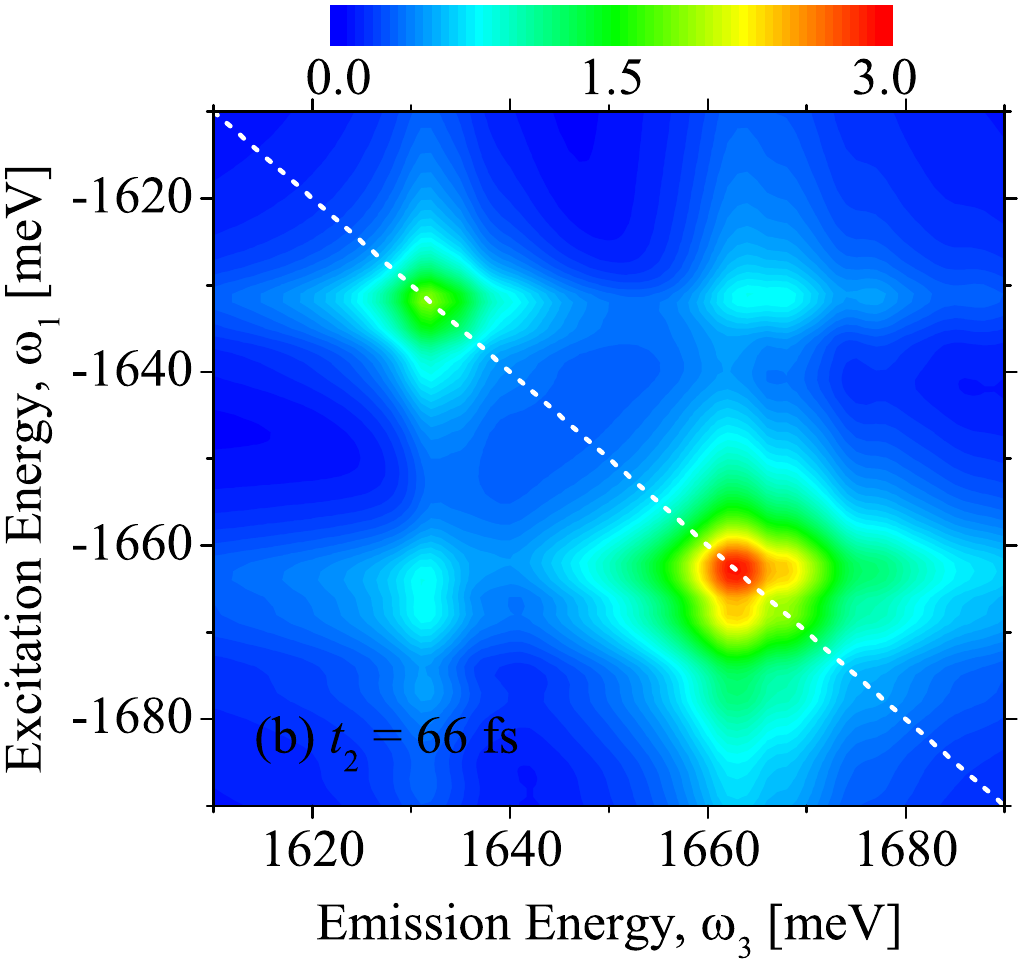}\includegraphics[width=0.33\textwidth]{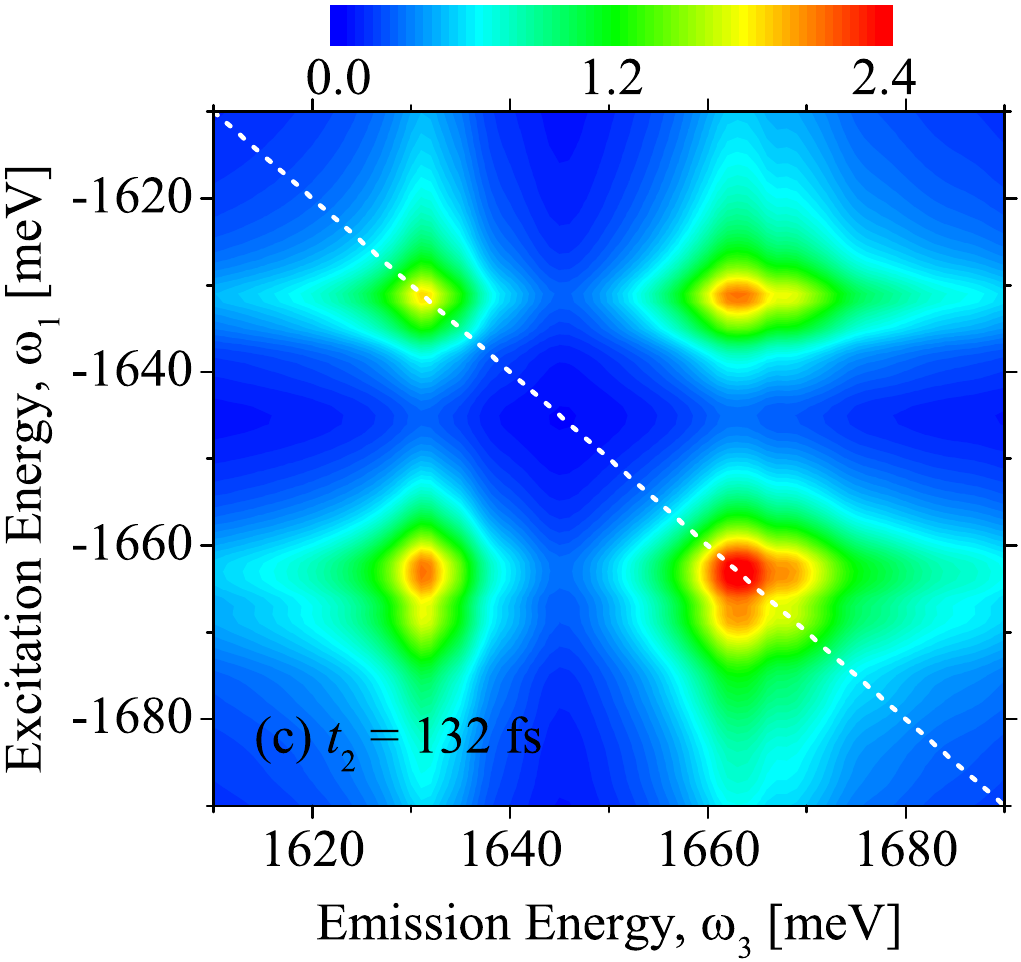}\caption{\label{fig:fig3_2DCSt2} The simulated rephasing 2D coherent spectra
(amplitude) at various mixing time decays $t_{2}$, to be compared
with the experimental data in Fig. 2 of Ref. \citep{Hao2016NanoLett}.
As in the experiment, the exciton ($X$) and trion ($T$) peaks appear
on the diagonal dashed line. The higher ($HCP$) and lower ($LCP$)
off-diagonal crosspeaks oscillate as a function of $t_{2}$, revealing
quantum coherence between excitons and trions. The red color illustrates
the maximum amplitude, as indicated in the colormap above each subplot.
The electron Fermi energy is set to be $\varepsilon_{F}=11.8$ meV.}
\end{figure*}

Let us now fix the parameters $t_{c}$, $t_{d}$, and $U$ in the
polaron model Hamiltonian, according to the recent experimental data
on monolayer MoSe$_{2}$ \citep{Hao2016NanoLett}. The electron mass
and hole mass in MoSe$_{2}$ are very similar, i.e., $m_{e}=m_{h}\simeq0.6m_{0}$
where $m_{0}$ is the free electron mass \citep{Wang2018}. Therefore,
the mass of the exciton should be two times larger, indicating $t_{d}=t_{c}/2$.
The hopping parameter $t_{c}$ might be estimated by the relation
$t_{c}=\hbar^{2}/(2m_{e}a^{2})$, where the lattice spacing of monolayer
MoSe$_{2}$ $a\sim3.2\textrm{Å}$ \citep{Tempelaar2019}. We find
then $t_{c}\sim620$ meV. However, we can not directly use this value,
considering the small electron density $n\sim10^{12}$ cm$^{-2}$
in the experiment \citep{Hao2016NanoLett}, which leads to $na^{2}=N/L^{2}\sim10^{-3}$.
Our numerical simulations have to be restricted to relatively small
lattice size, i.e., $L=16\sim N$, which has a density $na^{2}\sim0.06$.
To match the small electron density in the experiment, we need to
consider a coarse-grained model by artificially enlarging the lattice
spacing $a$ (i.e., making it 10 times larger). Thus, it seems reasonable
to take $t_{c}=10$ meV.

To determine the interaction strength $U<0$, we recall that the trion
binding energy is about 30 meV \citep{Hao2016NanoLett,Tempelaar2019}.
In our polaron model, this binding energy corresponds to the difference
between the repulsive polaron energy and the attractive polaron energy.
By performing numerical calculations with varying $U$ at a given
Fermi energy $\varepsilon_{F}=11.8$ meV (which corresponds to $N=24$
at the lattice size $L=16$), we find that $U=-6t_{c}$ reproduces
the observed trion binding energy, as reported in Fig. \ref{fig:fig2_akw}.
There, the simulated polaron spectral function has been rigidly shifted
by an amount $\omega_{X}=1662$ meV, so the repulsive polaron peak
lies at about 1663 meV. The photoluminescence spectrum of monolayer
MoSe$_{2}$ observed in the experiment \citep{Hao2016NanoLett} is
then qualitatively reproduced, by using our polaron model.

As a brief summary of the parameters to be used, throughout the work
we will use $t_{c}=10$ meV, $t_{d}=t_{c}/2=5$ meV and $U=-6t_{c}=-60$
meV. The lattice size is fixed to $L=16$. We tune the Fermi energy
$\varepsilon_{F}\simeq4\pi Nt_{c}/L^{2}$ by changing the number of
electrons $N$. To compare with the experimental 2DCS data \citep{Hao2016NanoLett},
the excitation energy $\omega_{1}$ and the emission energy $\omega_{3}$
will be shifted by $\omega_{X}=1662$ meV. As we use a finite-size
square lattice, the level spacing in the single-particle dispersion
relation is about $\delta=4t_{c}/L$. We will use $\delta$ to replace
the infinitesimal $0^{+}$ and to eliminate the discreteness in single-particle
energy levels. Finally, we would like to emphasize that, in our numerical
simulations, we do not include any phenomenological parameters such
as decoherence rates, which are often used to qualitatively understand
the experimental data \citep{Hao2016NanoLett,Tempelaar2019}.

\subsection{Quantum beats at the two crosspeaks}

In Fig. \ref{fig:fig3_2DCSt2}, we present the simulated rephasing
2D coherent spectra $\left|\mathcal{S}(\omega_{1},t_{2},\omega_{3})\right|$
at three mixing time decays $t_{2}=0$ (a), $t_{2}=66$ fs (b) and
$t_{2}=132$ fs (c). Although the electron density in the experiment
is unknown, we believe $\varepsilon_{F}=11.8$ meV, which corresponds
to the electron number $N=24$, could be a reasonable choice. The
three time delays are selected according to the measurements in Fig.
2(a)-(c) of Ref. \citep{Hao2016NanoLett}, so we can make an one-to-one
comparison. 

We find clearly the exciton ($X$) and trion ($T$) peaks along the
diagonal direction (see the dashed lines), as in the experiment. Furthermore,
two off-diagonal crosspeaks, labelled as $HCP$ and $LCP$, are fairly
evident. Their brightness oscillates with the mixing time delay $t_{2}$
as experimentally observed, revealing the coherent coupling between
excitons and trions. 

All those intriguing features can be understood from the simple expression
Eq. (\ref{eq:3rdResponseS}). At zero mixing time $t_{2}=0$, Eq.
(\ref{eq:3rdResponseS}) precisely predicts the existence of two diagonal
peaks at the attractive polaron (trion) energy $E_{T}=\mathcal{E}^{(n=0)}$
and at the repulsive polaron (exciton) energy $E_{X}$, respectively,
with strengths given by the residues $Z_{T}=\phi_{0}^{(0)*}\phi_{0}^{(0)}$
and $Z_{X}\sim1-Z_{T}$. The expression also predicts the two off-diagonal
crosspeaks at $(-\omega_{1},\omega_{3})=(E_{T},E_{X})$ and $(E_{X},E_{T})$,
with strength $\sqrt{Z_{T}Z_{X}}$. At nonzero mixing time $t_{2}\neq0$,
the quantum beats at the $HCP$ and $LCP$ crosspeaks can be easily
attributed to the term $e^{-i\left(\mathcal{E}_{n}-\mathcal{E}_{m}\right)T}$,
which gives rise to quantum oscillations with periodicity $2\pi/\left|E_{X}-E_{T}\right|$.

Let us now have a close comparison into the details. There is an apparent
quantitative discrepancy between theory and experiment on the crosspeak
brightness. Our prediction of the crosspeak strength $\sqrt{Z_{T}Z_{X}}$
means that at the zero mixing time $t_{2}=0$ the crosspeak brightness
should lie between those of the two diagonal peaks, as shown in Fig.
\ref{fig:fig3_2DCSt2}(a). However, this is not observed in the experiment
\citep{Hao2016NanoLett}. Experimentally, the crosspeaks are always
darker than the two main diagonal peaks, indicating the possibility
of some decoherence channels (i.e., the phonon-assisted up-conversion
and down-conversion processes as experimentally observed \citep{Hao2016NanoLett}). 

Apart from this discrepancy, we find a remarkable agreement between
theory and experiment on other details, upon changing the mixing time
$t_{2}$. As in the experiment, in each subplot of Fig. \ref{fig:fig3_2DCSt2},
the color scale is normalized to the highest peak (i.e., the exciton
peak $X$) in the spectra. For the trion peak $T$, we can see that
its \emph{relative} brightness is highest at $t_{2}=132$ fs and then
is a bit weaker at $t_{2}=0$ fs and $66$ fs. This subtle change
is precisely observed in the experiment \citep{Hao2016NanoLett}.
At the two crosspeaks, their relative brightness is similar at $t_{2}=0$
fs and 132 fs, which is also experimentally observed \citep{Hao2016NanoLett}. 

\begin{figure}
\centering{}\includegraphics[width=0.45\textwidth]{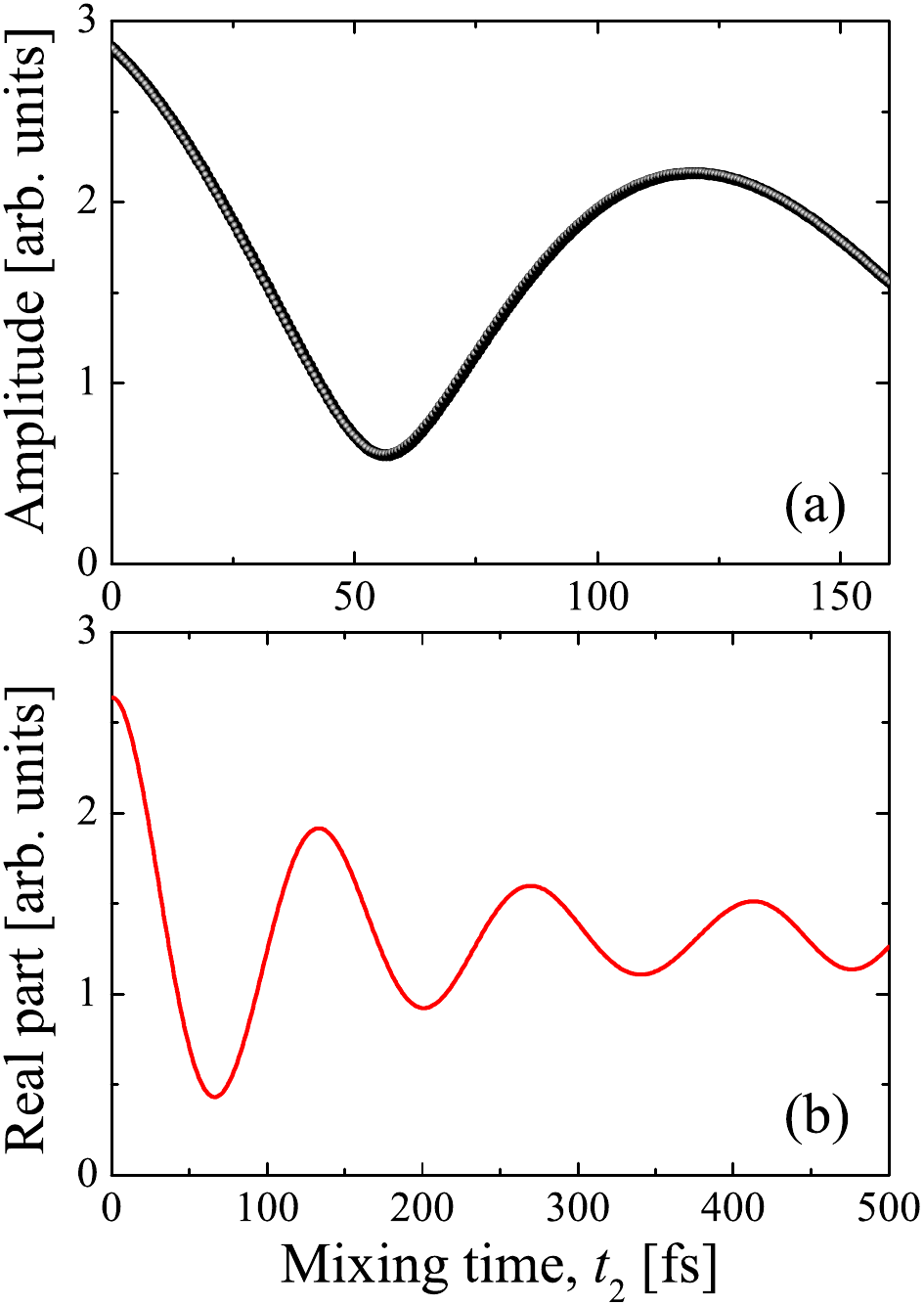}\caption{\label{fig:fig4_crosspeak} The simulated amplitude (a) and real part
(b) of the rephasing 2D signal at the crosspeaks as a function of
the mixing time delays $t_{2}$. The upper plot (a) is to be compared
with Fig. 3(b) of the experiment \citep{Hao2016NanoLett}. We take
the electron Fermi energy $\varepsilon_{F}=11.8$ meV.}
\end{figure}

In Fig. \ref{fig:fig4_crosspeak}, we report the simulated rephasing
2D signal at the crosspeaks as a function of the mixing time $t_{2}$.
The amplitude of the 2D signal in Fig. \ref{fig:fig4_crosspeak}(a)
should be contrasted with Fig. 3(b) of the experiment \citep{Hao2016NanoLett}.
Our simulation reproduces very well the quantum oscillations observed
in the experiment, with essentially the same periodicity. However,
we note that, in spite of the same periodicity the two oscillations
at $HCP$ and $LCP$ crosspeaks measured in the experiment are slightly
unsynchronized. Our theory always predicts the exactly same oscillation
at the two crosspeaks, as the predicted 2DCS response Eq. (\ref{eq:3rdResponseS})
is \emph{symmetric} upon switching the excitation and emission energies,
as we emphasized earlier.

On the other hand, the real part of the 2D signal in Fig. \ref{fig:fig4_crosspeak}(b)
might be compared with Fig. 3(b) of the pioneering work by Tempelaar
and Berkelbach \citep{Tempelaar2019}. There is an excellent agreement
in curve shape and periodicity. The only difference is that our 2D
signal never decays to zero. This is simply due to the ground-state
bleaching (GSB) process illustrated in Fig. \ref{fig:fig1_sketch}(c),
which has not taken into account in Ref. \citep{Tempelaar2019} but
gives an important $t_{2}$-independent 2D signal.

\begin{figure*}
\begin{centering}
\includegraphics[width=0.4\textwidth]{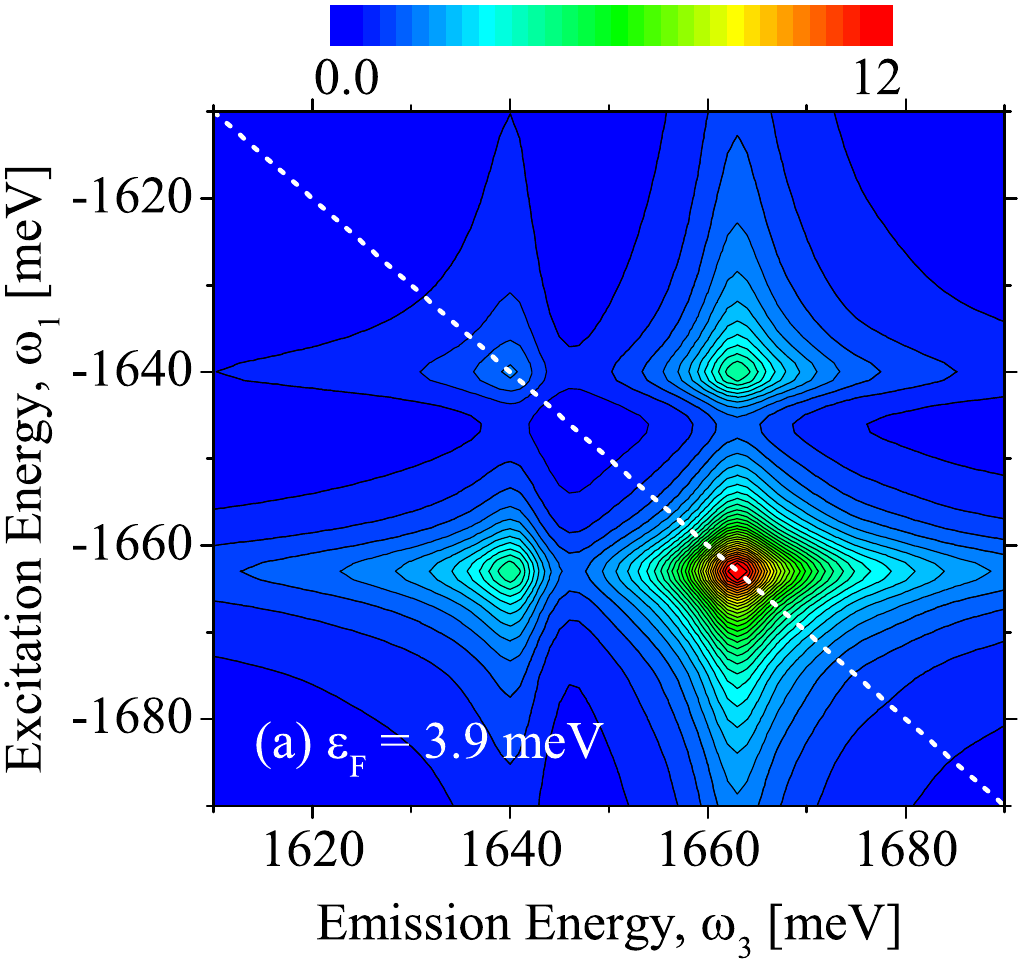}\includegraphics[width=0.4\textwidth]{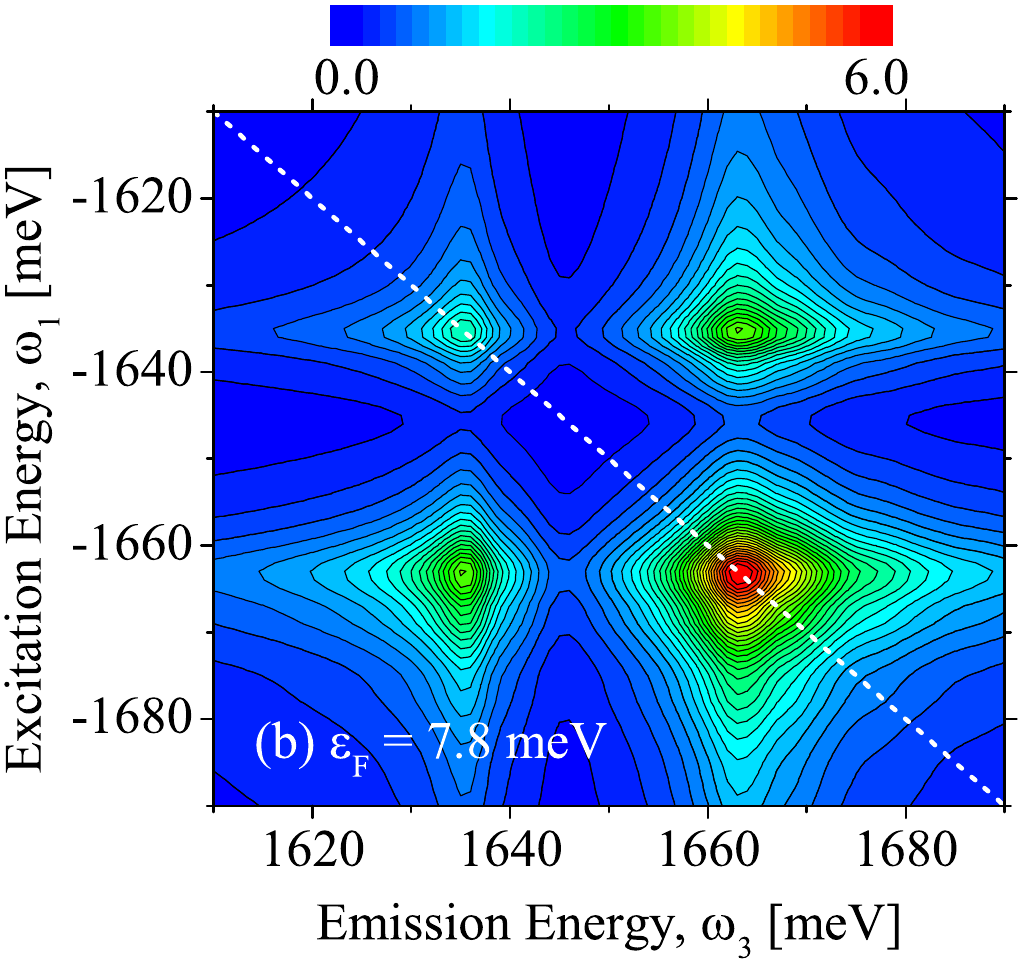}
\par\end{centering}
\centering{}\includegraphics[width=0.4\textwidth]{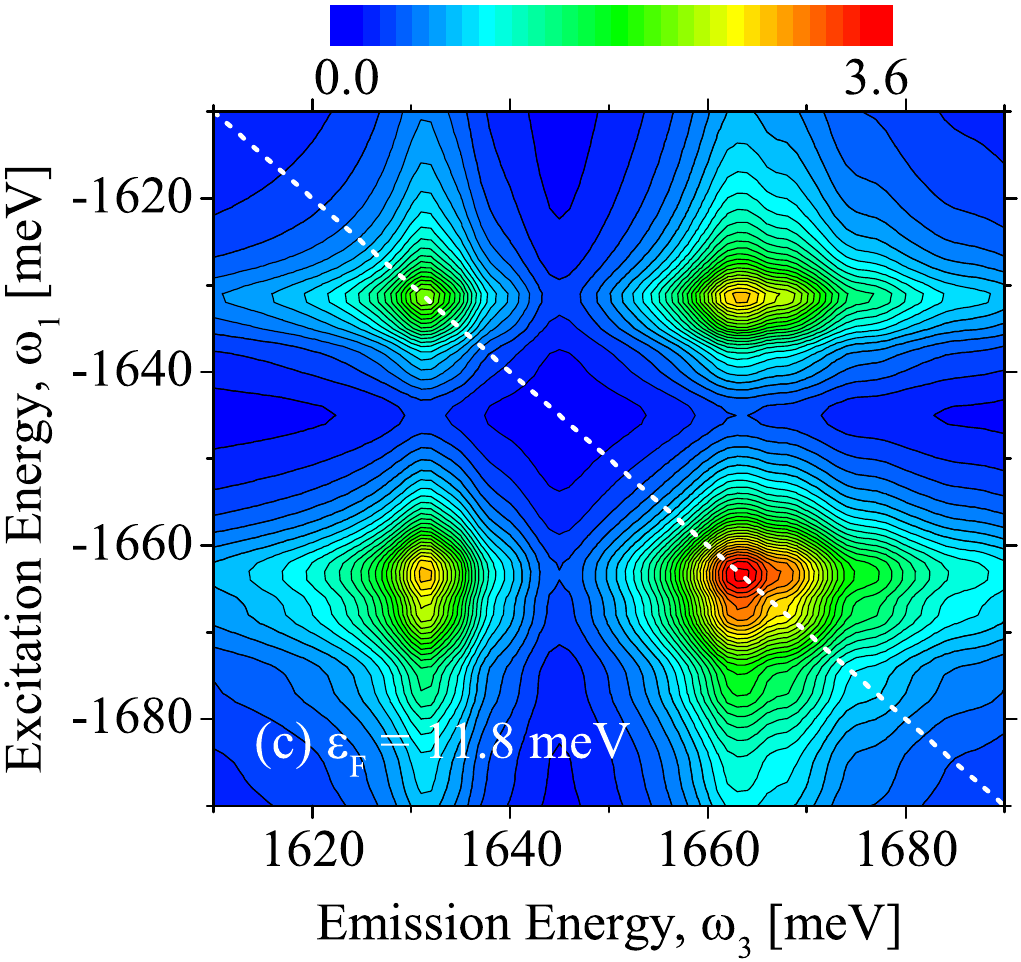}\includegraphics[width=0.4\textwidth]{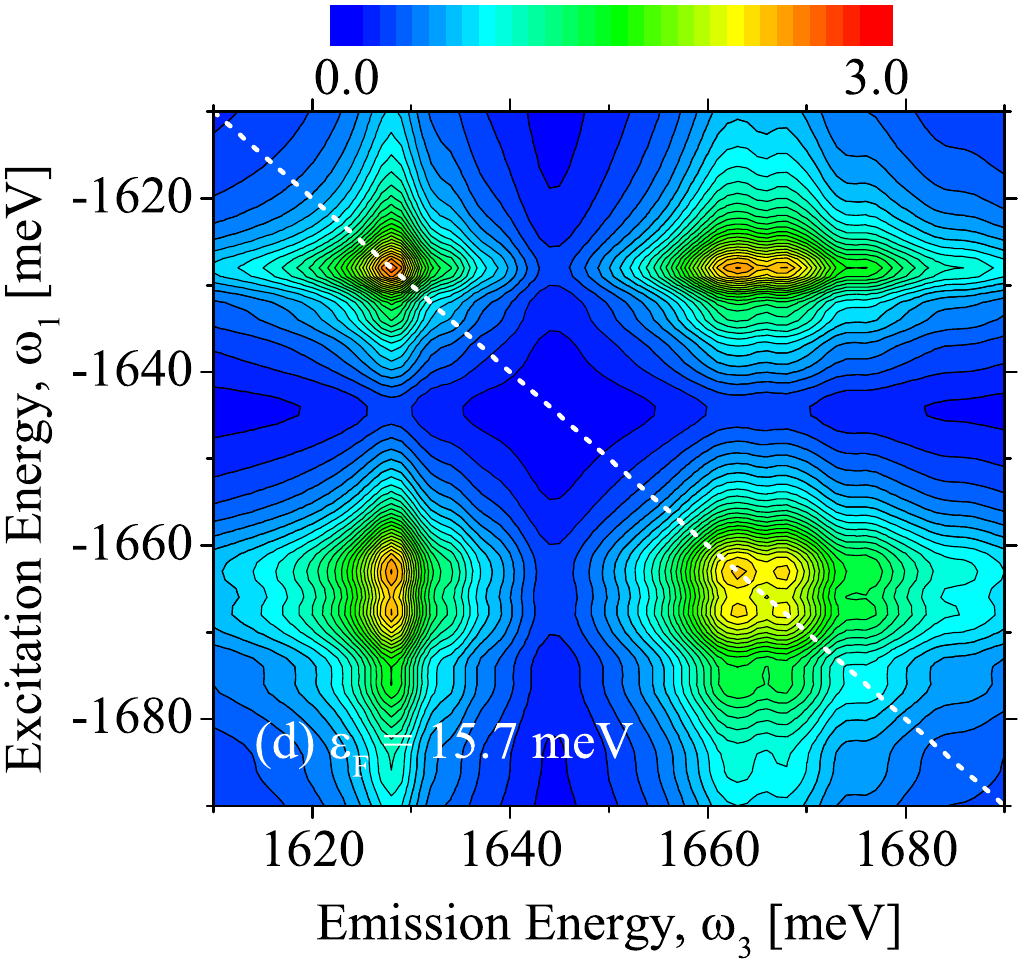}\caption{\label{fig:fig5_2DCS0} The simulated rephasing 2D coherent spectra
(amplitude) with increasing electron density or Fermi energy, at zero
mixing time delay $t_{2}=0$. The red color illustrates the maximum
amplitude, as indicated in the colormap above each subplot.}
\end{figure*}

\subsection{Dependence on the electron density}

We finally consider the dependence of the rephasing 2D signal on the
electron density or the electron Fermi energy at the mixing time $t_{2}=0$.
As shown in Fig. \ref{fig:fig5_2DCS0}, as the density increases,
the attractive polaron (or trions) peak acquires larger brightness
(i.e., oscillation strength) and has a red-shift in energy with respect
to the repulsive polaron peak (excitons). This observation agrees
well the existing measurements \citep{Wang2018} on the reflection
(absorption) spectra of 2D materials and also the relevant theoretical
explanations \citep{Tempelaar2019}. The brightness of the two off-diagonal
crosspeaks also increases with increasing electron density. This theoretical
prediction could be examined in future 2DCS measurements with a controllable
electron density.

\section{Conclusions and outlooks}

In summary, we have investigated the 2D coherent spectroscopy of excitons
and trions in monolayer transition metal dichalcogenides, by using
a many-body Fermi polaron model with mobile exciton. Our investigation
complements the previous pioneering studies based on either few-body
calculations \citep{Tempelaar2019} or the exact solutions in immobile
heavy exciton limit \citep{Lindoy2022}. We have derived a simple
expression for the 2D coherent spectroscopy, which is applicable to
the limit of a single exciton. By performing numerical simulations
without any phenomenological parameters, we have found that this simple
expression captures the essential features of the observed 2D coherent
spectroscopy of monolayer MoSe$_{2}$ and yields an excellent agreement
with the experiment \citep{Hao2016NanoLett}. 

There are residual discrepancies at the quantitative level. For example,
the predicted crosspeak relative brightness is higher than what has
been observed \citep{Hao2016NanoLett} and our theory is unable to
explain the slightly unsynchronized quantum beats at different crosspeaks
in the experiment \citep{Hao2016NanoLett}. Presumably, we feel that
the polaron-polaron interaction (that we have neglected in our treatment)
could be one of the sources for these discrepancies. To take into
account the polaron-polaron interaction, we need to consider at least
two excitons. The third rephasing process of excited-state absorption
(i.e., the $R_{1}^{*}(t_{1},t_{2},t_{3})$ term) then would make an
important contribution.

On the other hand, experimentally, the different implementation of
the polarization of the laser pulses (such as the pathway $\sigma^{+}\sigma^{-}\sigma^{+}\sigma^{-}$)
can be used to create bi-excitons \citep{Hao2017} and bi-polarons
\citep{Muir2022} in the many-body dynamics of monolayer 2D materials.
It would be interesting to extend our theoretical framework to explain
the experimental data in those situations.
\begin{acknowledgments}
This research was supported by the Australian Research Council's (ARC)
Discovery Program, Grants No. DE180100592 and No. DP190100815 (J.W.),
and Grant No. DP180102018 (X.-J.L).
\end{acknowledgments}


\begin{thebibliography}{10}
\bibitem{Novoselov2005}K. S. Novoselov, D. Jiang, F. Schedin, T.
J. Booth, V. V. Khotkevich, S. V. Morozov, and A. K. Geim, Two-dimensional
atomic crystals, Proc. Natl. Acad. Sci. U.S.A.\textbf{ 102}, 10451
(2005). 

\bibitem{Wang2018}G. Wang, A. Chernikov, M. M. Glazov, T. F. Heinz,
X. Marie, T. Amand, and B. Urbaszek, Colloquium: Excitons in atomically
thin transition metal dichalcogenides, Rev. Mod. Phys. \textbf{90},
021001 (2018).

\bibitem{Berkelbach2018}T. C. Berkelbach and D. R. Reichman, Optical
and Excitonic Properties of Atomically Thin Transition-Metal Dichalcogenides,
Annu. Rev. Condens. Matter Phys. \textbf{9}, 379 (2018).

\bibitem{Mak2010}K. F. Mak, C. Lee, J. Hone, J. Shan, and T. F. Heinz,
Atomically Thin MoS$_{2}$: A New Direct-Gap Semiconductor, Phys.
Rev. Lett. \textbf{105}, 136805 (2010).

\bibitem{Gutierrez2013}H. R. Gutiérrez, N. Perea-Lopez, A. L. Elías,
A. Berkdemir, B. Wang, R. Lv, F. Lopez-Urías, V. H. Crespi, H. Terrones,
and M. Terrones, Extraordinary room-temperature photoluminescence
in triangular WS$_{2}$ monolayers, Nano Lett. \textbf{13}, 3447 (2013).

\bibitem{Ahmed2020}S. Ahmed, X. Jiang, F. Zhang, and H. Zhang, Pump--probe
micro-spectroscopy and 2D materials, J. Phys. D: Appl. Phys. \textbf{53},
473001 (2020).

\bibitem{Tan2020}L. B. Tan, O. Cotlet, A. Bergschneider, R. Schmidt,
P. Back, Y. Shimazaki, M. Kroner, and A. \.{I}mamo\u{g}lu, Interacting
Polaron-Polaritons, Phys. Rev. X \textbf{10}, 021011 (2020).

\bibitem{Hao2016NatPhys}K. Hao, G. Moody, F. Wu, C. K. Dass, L. Xu,
C.-H. Chen, L. Sun, M.-Y. Li, L.-J. Li, A. H. MacDonald, and X. Li,
Direct measurement of exciton valley coherence in monolayer WSe$_{2}$,
Nat. Phys. \textbf{12}, 677 (2016).

\bibitem{Jonas2003}D. Jonas, Two-Dimensional Fermtosecond Spectroscopy,
Ann. Rev. Phys. Chem. \textbf{54}, 425 (2003).

\bibitem{Li2006}X. Li, T. Zhang, C. N. Borca, and S. T. Cundiff,
Many-Body Interactions in Semiconductors Probed by Optical Two-Dimensional
Fourier Transform Spectroscopy, Phys. Rev. Lett. \textbf{96}, 057406
(2006).

\bibitem{Cho2008}M. Cho, Coherent Two-Dimensional Optical Spectroscopy,
Chem. Rev. \textbf{108}, 1331 (2008).

\bibitem{Hao2016NanoLett}K. Hao, L. Xu, P. Nagler, A. Singh, K. Tran,
C. K. Dass, C. Schuller, T. Korn, X. Li, and G. Moody, Coherent and
incoherent coupling dynamics between neutral and charged excitons
in monolayer MoSe$_{2}$, Nano Lett. \textbf{16}, 5109 (2016).

\bibitem{Hao2017}K. Hao, J. F. Specht, P. Nagler, L. Xu, K. Tran,
A. Singh, C. K. Dass, C. Schuller, T. Korn, M. Richter, A. Knorr,
X. Li, and G. Moody, Neutral and charged inter-valley biexcitons in
monolayer MoSe$_{2}$, Nat. Commun. \textbf{8}, 15552 (2017).

\bibitem{Muir2022}J. B. Muir, J. Levinsen, S. K. Earl, M. A. Conway,
J. H. Cole, M. Wurdack, R. Mishra, D. J. Ing, E. Estrecho, Y. Lu,
D. K. Efimkin, J. O. Tollerud, E. A. Ostrovskaya, M. M. Parish, and
J. A. Davis, Exciton-polaron interactions in monolayer WS$_{2}$,
arXiv:2206.12007 (2022).

\bibitem{Tempelaar2019}R. Tempelaar and T. C. Berkelbach, Many-body
simulation of two-dimensional electronic spectroscopy of excitons
and trions in monolayer transition metal dichalcogenides, Nat. Commun.
\textbf{10}, 3419 (2019).

\bibitem{Lindoy2022}L. P. Lindoy, Y.-W. Chang, and D. R. Reichman,
Two-dimensional spectroscopy of two-dimensional materials, arXiv:2206.01799
(2022).

\bibitem{Sidler2017}M. Sidler, P. Back, O. Cotlet, A. Srivastava,
T. Fink, M. Kroner, E. Demler, and A. Imamoglu, Fermi polaron-polaritons
in chargetunable atomically thin semiconductors, Nat. Phys. \textbf{1}3,
255 (2017).

\bibitem{Efimkin2017}D. K. Efimkin and A. H. MacDonald, Many-body
theory of trion absorption features in two-dimensional semiconductors,
Phys. Rev. B \textbf{95}, 035417 (2017).

\bibitem{Massignan2014}P. Massignan, M. Zaccanti, and G. M. Bruun,
Polarons, dressed molecules and itinerant ferromagnetism in ultracold
Fermi gases, Rep. Prog. Phys. \textbf{77}, 034401 (2014).

\bibitem{Schmidt2018}R. Schmidt, M. Knap, D. A. Ivanov, J.-S. You,
M. Cetina, and E. Demler, Universal many-body response of heavy impurities
coupled to a Fermi sea: a review of recent progress, Rep. Prog. Phys.
\textbf{81}, 024401 (2018).

\bibitem{Wang2022PRL}J. Wang, X.-J. Liu, and H. Hu, Exact Quasiparticle
Properties of a Heavy Polaron in BCS Fermi Superfluids, Phys. Rev.
Lett. \textbf{128}, 175301 (2022).

\bibitem{Wang2022PRA}J. Wang, X.-J. Liu, and H. Hu, Heavy polarons
in ultracold atomic Fermi superfluids at the BEC-BCS crossover: Formalism
and applications, Phys. Rev. A \textbf{105}, 043320 (2022).

\bibitem{Mahan1967}G. D. Mahan, Excitons in Metals: Infinite Hole
Mass, Phys. Rev. \textbf{163}, 612 (1967).

\bibitem{Nozieres1969}P. Nozières and C. T. De Dominicis, Singularities
in the X-Ray Absorption and Emission of Metals. III. One- Body Theory
Exact Solution, Phys. Rev. \textbf{178}, 1097 (1969).

\bibitem{Mahan2008}G. D. Mahan, \textit{Many-Particle Physics} (Springer
India, 3rd edition, January 1, 2008), Chapter 9.

\bibitem{Anderson1967}P. W. Anderson, Infrared Catastrophe in Fermi
Gases with Local Scattering Potentials, Phys. Rev. Lett. \textbf{18},
1049 (1967).

\bibitem{Knap2012}M. Knap, A. Shashi, Y. Nishida, A. Imambekov, D.
A. Abanin, and E. Demler, Time-Dependent Impurity in Ultracold Fermions:
Orthogonality Catastrophe and Beyond, Phys. Rev. X \textbf{2}, 041020
(2012).

\bibitem{Wang2022arXiv1}J. Wang, Multidimensional Spectroscopy of
Time-Dependent Impurities in Ultracold Fermions, arXiv:2207.10501
(2022).

\bibitem{Wang2022arXiv2}J. Wang, H. Hu, and X.-J. Liu, Two-dimensional
spectroscopic diagnosis of quantum coherence in Fermi polarons, arXiv:2207.14509
(2022).

\bibitem{Mukamel1995}S. Mukamel, \textit{Principles of nonlinear
optical spectroscopy} (Oxford University Press, 1995).

\bibitem{Zhang2008}Tianhao Zhang, \textit{Optical Two-Dimensional
Fourier Transform Spectroscopy of Semiconductors} (PhD thesis, University
of Colorado, 2008), Chapter 3.

\bibitem{Chevy2006}F. Chevy, Universal phase diagram of a strongly
interacting Fermi gas with unbalanced spin populations, Phys. Rev.
A \textbf{74}, 063628 (2006).
\end{thebibliography}
\end{document}